# Moving towards nano-TCAD through multi-million atom quantum dot simulations matching experimental data

Muhammad Usman, Hoon Ryu, Insoo Woo, David S. Ebert, and Gerhard Klimeck (Senior Member IEEE)

*Abstract*—Low-loss optical communication requires light sources at 1.5um wavelengths. Experiments showed without much theoretical guidance that InAs/GaAs quantum dots (QDs) may be tuned to such wavelengths by adjusting the *In* fraction in an $In_xGa_{1-x}As$ strain-reducing capping layer (SRCL). In this work systematic multi-million atom electronic structure calculations qualitatively and quantitatively explain for the first time available experimental data. The NEMO 3-D simulations treat strain in a 15 million atom system and electronic structure in a subset of ~9 million atoms using the experimentally given nominal geometries and without any further parameter adjustments the simulations match the non-linear behavior of experimental data very closely. With the match to experimental data and the availability of internal model quantities significant insight can be gained through mapping to reduced order models and their relative importance. We can also demonstrate that starting from simple models has in the past led to the wrong conclusions. The critical new insight presented here is that the QD changes its shape. The quantitative simulation agreement with experiment without any material or geometry parameter adjustment in a general atomistic tool leads us to believe that the era of nano Technology Computer Aided Design (nano-TCAD) is approaching. NEMO 3-D will be released on nanoHUB.org where the community can duplicate and expand on the results presented here through interactive simulations.

Manuscript received July 13, 2008. This work was carried out in part at the Jet Propulsion Laboratory (JPL), California Institute of Technology, under a contract with the National Aeronautics and Space Administration (NASA). Muhammad Usman is funded through Fulbright USAID (Grant ID # 15054783). nanoHUB.org computational resources operated by the Network for Computational Nanotechnology (NCN) funded by the National Science Foundation were used in this work.

Muhammad Usman is with the Network for Computational Nanotechnology, Electrical Engineering Department, Purdue University West Lafayette, Indiana 47907 USA (Corresponding author: Phone: 765-418-9489; fax: 765-494-6441; e-mail: usman@purdue.edu ).

Gerhard Klimeck is with the Network for Computational Nanotechnology, Electrical Engineering Department, Purdue University West Lafayette, Indiana USA (email: gekco@purdue.edu )

Hoon Ryu, Insoo Woo, and David S. Ebert are with the Electrical Engineering Department, Purdue University West Lafayette, Indiana 47907 USA (email: {ryu2@purdue.edu, iwoo@purdue.edu, })

## I. INTRODUCTION AND METHOD

Quantum dots grown by self-assembly are typically constructed by 50,000 to 10,000,000 atoms which confine a small, countable number of extra electrons or holes in a space that is comparable in size to the electron wavelength. Under such conditions quantum dots can be interpreted as artificial atoms which can be custom tailored to new functionality. As such these structures have attracted significant experimental and theoretical attention in the field of nanoscience [1]. The new and tunable optical and electrical properties of these artificial atoms have been proposed in a variety of different fields, for example in communication and computing systems, medical and quantum computing applications. Predictive and quantitative modeling and simulation may help to narrow down the vast design space to a range that is experimentally affordable and move this part of nanoscience to nanotechnology.

After 10 years of development we demonstrate here the capabilities of the Nanoelectronic Modeling (NEMO 3-D) toolkit which can now quantitatively model complex sequence of multiple experiments without any model parameter adjustments. Input to the calculations are the previously published atomistic material parameters and the experimental device geometries and alloy concentrations. The resulting numerical simulation domain includes around 15 million atoms to represent the realistic device. Quantitative agreement of the simulations with a non-linear behavior across a sequence of different experimental devices is found. The internal model variables such as strain, atom positions, and disorder can subsequently be used to gain physical insight and intuition into the competing processes observed experimentally, by mapping into simplified models. It is, however, critical to start from the full atomistic model in order to include the important effects initially, rather than trying to access the relative importance of the simple models. It is shown that starting from the simple models has, in the past, led to wrong conclusions.

Figure 1 shows two views of the atomistically represented quantum dot (QD) system. An InAs dome shaped QD is grown on top of an InAs wetting layer and covered with an $In_{0.4}Ga_{0.6}As$ (Fig1a) or $In_{0.4}Al_{0.6}As$ (fig1b) capping layer, followed by a GaAs cap. The regularly distributed 'As' atoms are not shown for clarity. The atomic disorder in the alloyed capping layer as well as the InAs QD interface is evident. The *In* fraction $x$ in the $In_xGa_{1-x}As$ cap layer has been varied in experiments to tune the optical emission spectrum without much theoretical guidance. The NEMO 3-D simulation is based on this atomistic device representation and intrinsically incorporates the disordered atom placement, and our model can explain the observed experiment non-linear behavior without any parameter adjustments.

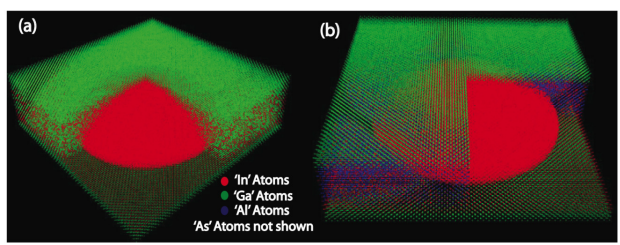

Fig. 1. Two views of the atomistically represented quantum dot device systems. An InAs dome shaped quantum dot is grown on top of an InAs wetting layer and covered with a (a) InGaAs (b) InAlAs capping layer, followed by a GaAs cap. The 'As' atoms are not shown for clarity purpose. The atomic disorder in the alloyed capping layer as well as the InAs quantum dot interface is evident.

Much interest in recent years has been focused on GaAs-based optical devices used for optical-fiber based communication systems at longer wavelengths (1.3μm – 1.5μm). Several experimental groups have tried to achieve emissions at wavelengths above 1.3μm either by using GaInNAs quantum dots in a GaAs matrix [2] or by embedding InAs QDs in a $In_xGa_{1-x}As$ Strain-Reducing Capping Layer (SRCL) [3-8]. To date, there is not much theoretical analysis of SRCL capped systems available in the literature and the dependence of emitted spectra wavelength on the *In* fraction and the thickness of the SRCL has not been studied in detail. The effect of a graded interface between the QD and the surrounding buffer as well as QD size variations has not been investigated yet either. Ref. 13 used an eight band k•p method to explain the observed red shift without giving any quantitative contribution of QD size change and reduction of barrier height on SRCL side. A single band effective mass approximation is used in Ref. 5. That work claims that the reduction of confinement barrier height on the SRCL side is a major source of observed red shift. In this paper, we qualitatively and quantitatively explain the experimentally observed red shift of the emission spectra in terms of the strain relaxation in the growth direction, the change in the volume of the QDs and the reduction of the barrier height on the SRCL side in contradiction to the reduced order models used before.

Our atomistic tight binding model clarifies the physical picture and shows that this barrier height lowering makes only a very minor contribution to the red shift, in contrast to the explanation of the continuum models. A critical insight needed to solve this problem is the atomistic representation of the system rather than the continuum picture and the subsequent consequences of the atomistic representation on the electronic structure.

A comprehensive analysis of the strain relaxation, the QD size changes and the barrier reduction on the SRCL side is presented for strain-reduced QDs using the NanoElectronic MOdeling package NEMO 3-D [9, 27]. NEMO 3-D can calculate strain and electronic structure for realistically sized systems as large as 52 million atoms [9] which corresponds to a simulation domain of $(101nm)^3$. Strain is calculated using an atomistic Valence Force Field (VFF) method [11] and the electronic structure using a twenty band $sp^3d^5s^*$ nearest neighbor empirical tight binding model [10]. The tight binding parameters are fit to reproduce the bulk properties of GaAs, InAs, AlAs, Si, and Ge with respect to room temperature band edges, effective masses, hydrostatic/bi-axial strain behavior, and relative band edges [10, 29, 34-36] using a global minimization procedure based on a genetic algorithm [28] and analytical insight [29]. The strain and electronic structure properties of alloys are faithfully reproduced through an explicit disordered atomistic representation [27, 29] rather than an averaged potential representation. The explicit alloy representation also affords the ability to model device-to-device fluctuations, which are critical in modern deca-nano devices. The bulk-based atom-to-atom interactions are transferred into nano-scale devices where no significant bond charge redistribution or bond breaking is expected and strain is typically limited to around 8%. To material scientists and *ab-initio* theorists this model might appear to have a very limited range of validity, but it basically covers all standard semiconductors relevant for realistic devices where bonds and geometries are stable and contain tens of thousands to tens of millions of atoms. In fact one might argue that a device in which bonds are established and dissolved result in undesirable electronic structure fluctuations for the realistic devices. For realistic semi-conducting nano-scale systems our tight binding approach, employed in NEMO 1-D, and NEMO 3-D, has been validated experimentally through 1) high bias, high current, quantitative resonant tunneling diode modeling [30] (unprecedented match to a suite of experimental data of room temperature quantum effect devices), 2) photoluminescence in InAs nanoparticles [31], 3) modeling of the Stark effect of single P impurities in Si [32], (demonstrating the need to include an atomistic representation rather than continuum effective mass models), 4) distinguishing P and As impurities in ultra-scaled FinFET devices [40, 42] and 5) the valley splitting in miscut Si quantum wells on SiGe substrate [33] (demonstrating importance of the atomistic wafer step disorder and alloy disorder in SiGe). In this work NEMO 3-D is used as an analysis and design tool without any material parameter tuning. Semi-analytical, reduced-order models are used to provide insight into the meaning and validity of the numerical results.

Understanding the effects of strain in an InAs/GaAs system is of critical importance [12, 13, 18]. The introduction of a SRCL modifies the strain distributions significantly, and the hydrostatic and the biaxial strain relaxations for varying *In* fractions in the SRCL surrounding the InAs QD need to be studied. It is shown that strain relaxation inside the InAs QD has a significant contribution in the observed red shift of emission spectra for small *In* concentrations ($x<0.2$). For large *In* fractions ($x>0.2$), another physical modification in the system begins to play a major role in the additional red shift. The physical shape of the QD changes where a decrease in the base length and a relatively larger increase in the height of the InAs QD modify the quantum confinement and therefore change the confinement energy. The two strain induced effects of band edge changes and shape changes are mixed non-linearly due to differing In-As and Ga-As bond-lengths of the atomistically disordered InGaAs buffer system. NEMO 3-D includes all these effects due to its fundamental atomistic material representations and faithfully reproduces the experimentally observed [3] non-linear behavior of the emission wavelength as a function of *In* concentration. Simple effective mass models of this system predict [5] that the introduction of the InGaAs buffer reduces the electronic confinement potential relative to the GaAs buffer. Our atomistic simulations show that this effect does not take place by direct comparison to an InAlAs buffer, which has the same strain, but different electronic structure behavior as the InGaAs buffer.

Significant nonlinearity in the biaxial strain components can be attributed to the bi-modal In-As and Ga-As bond distributions in InGaAs which can only be captured by an atomistic representation. Additional case studies show that the non-linearity in the optical wavelength tuning profile is stable relative to experimentally unavoidable effects such as interface inter-diffusion (interface softness) and quantum dot size variations.

NEMO 3-D can be used to gain fundamental insight into the key physics and guide future experiments with modest computational expense and time at about 10 hours on 60 CPUs per simulation. Efforts are under way to enable anyone in the community to perform these NEMO 3-D simulations on nanoHUB.org [24] which already hosts over 120 nano simulation tools, including an educational version [43] of NEMO 3-D. A full capability NEMO 3-D release is planned for the fall of 2008. The educational NEMO 3-D version entitled "Quantum Dot lab" runs in a few seconds and has been used by over 1,600 users who ran over 14,000 simulations.

II. SYSTEM SIMULATED AND MATCH TO EXPERIMENT

Figure 2(a) shows the schematic of the simulated system which consist of a lens shaped InAs quantum dot (QD) of 5nm height and 20nm base diameter on a 1ML InAs wetting layer. The QD is embedded in a $In_xGa_{1-x}As$ SRCL of height "$D$" nm. A large GaAs buffer (a total volume of $60\times60\times66$ nm$^3$) surrounds the whole structure containing 15.2 million atoms which is used for the atomistic strain calculation. Physical intuition tells us that the states of interest are confined in the region around the QD. Numerical experiments confirm that the electronic structure computation can be performed in a smaller region (50nm in the lateral dimension and 56nm in the growth direction consisting of 8.96 million atoms) around the InAs QD without any loss of accuracy, after the atom positions have been computed in the large domain, due to the long-range strain interaction. Two different values of *D* are considered: *D* = 5nm and *D* = 10nm, subsequently labeled D5, D10, respectively. These systems are the nominal topologies given in the experiments [3, 4]. The *In* fractions ($x$) in the SRCL take the values of 0 (no capping layer), 0.12, 0.18, 0.28, 0.4, and 0.45.

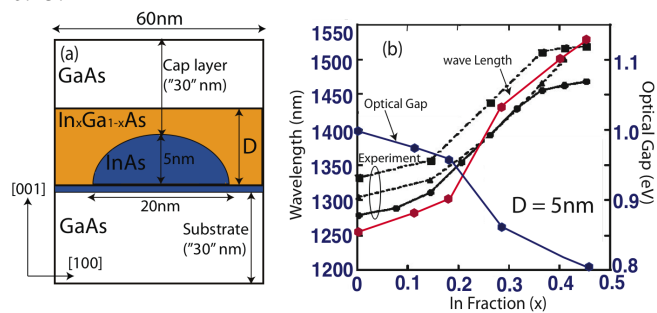

Fig. 2. (a) Schematic of the system simulated containing InAs quantum dot on 1ML InAs wetting layer. The dot is placed in $In_xGa_{1-x}As$ strain-reducing cap layer of thickness "*D*". The QD is dome shaped with 5nm height and 20nm diameter. The thickness *d* of $In_xGa_{1-x}As$ SRCL takes the values 5nm (*D5*) and 10nm (*D10*) (b) Optical band gap and the emission wavelength as a function of the *In* fraction ($x$) in SRCL in *D5*. The red curve is computed from NEMO 3-D simulations. The black curves are experimental values from Ref 3. The three black lines are for the same InGaAs compositions grown by low-pressure MOCVD using trimethylindium (TMI), trimethylgallium (TMG), triethylgallium (TEG), and tertiarybutylarsine (TBA) as the source materials at the total pressure of 76 Torr. Three samples of the same QDs are grown by changing the growth conditions of InAs QDs, and the source materials of the GaAs capping layer. The conditions of the TBA partial pressure, the growth rate of InAs QDs, and the source materials of GaAs capping layer are shown in Table 1 of Reference 3.The authors in Ref. 3 have reported mean diameter of QDs to be 20nm and mean height to be 5nm. We calculate an emission wavelength of 1.53um for experimental value of ~1.52um at x=0.45.

Figure 2(b) compares the optical emission wavelength observed in a complex experimental [3] (black curves) device sequence against NEMO 3-D simulations (red curve). The three experimental curves are obtained at the different growth conditions resulting in nominally identical dots of average diameter of 20nm and height of 5nm. The NEMO 3-D simulations take as input parameters the geometry and alloy compositions, as described in the experimental publications, without any other material parameter or geometry adjustments. Previously published material parameters [10] are used and no parameters other than the quantum dot size, wetting layer thickness and the alloy composition are entered. A surprisingly good agreement between the experimental results and the "out-of-box" NEMO 3-D calculations is observed. If this were production-level, conventional nanotechnology Computer Aided Design (nano-CAD) engineering work, one might conclude that NEMO 3-D is a tool that can be used to explore this design space quantitatively. We will soon release this code on nanoHUB.org [24].

The experiment and simulations show an interesting non-linear behavior whose origin is not clearly understood. The remainder of this paper reveals the intriguing details that lead to this complex nonlinear behavior and provides insight into the physical problem through analysis of internal model variables and derived simplified analytical models. The critical insights gained from this study are the following:

- Qualitative and quantitative explanation of physics:

    o Atomistic material representation includes bi-modal In-As and Ga-As bond-lengths
      ➔ non-linear strain distribution as a function of In concentration

    o Strain relaxation in growth direction
      ➔ red shift of emission spectra

    o Quantum dot shape changes (Aspect Ratio)
      ➔ red shift of emission spectra

    o Changes in electronic confinement of the InGaAs buffer barrier height
      ➔ negligible contribution to change in emission spectra

    o Atomistic interface detail
      ➔ softness of QD boundary causes a small blue shift.

    o QD size variations
      ➔ little effect on conclusions, can provide guidance to device design.

- Conceptual observations and conclusions

    o Details of atomistic strain and electronic calculations are important
      ➔ effective mass models and continuum theories lead to wrong conclusions.

    o NEMO 3-D electronic structure at room temperature matches experiments more closely than pseudo-potential method for the InGaAs system (confinement energies and bad gaps).

    o NEMO 3-D matches experimental InAs/GaAs Quantum Dot data out-of-box without any parameter adjustments, like NEMO 1-D 10 years earlier [41] for InP and GaAs-based resonant tunneling diodes
      ➔ nano-TCAD is approaching for quantum dot systems.

III – RESULTS AND DISCUSSIONS

A. *Strain relaxation impact*

Figure 3 shows the hydrostatic $\{\varepsilon_H = \varepsilon_{xx}+\varepsilon_{yy}+\varepsilon_{zz}\}$ and the biaxial strain $\{\varepsilon_B = \varepsilon_{xx}+\varepsilon_{yy}-2\varepsilon_{zz}\}$ components [15] for different *In* fractions and heights of the SRCL. Only the spatial regions around the central QD are shown for clarity. The actual device simulation domain is significantly larger (figure 2(a)) where the substrate and lateral strain have decayed. The hydrostatic strain is strictly negative everywhere, large only within the InAs QD and almost zero outside. The larger bond-length InAs QD is compressed by the smaller bond-length GaAs environment in the lateral direction. The peak value variation of the hydrostatic strain within the InAs QD in the [001] and [100] directions are shown at the bottom of the left column of figure 3. As the *In* fraction (*x*) increases from 0 to 0.4, the hydrostatic strain decreases and the peak value of the strain shifts from the top towards the center of the dot. This shift of the peak hydrostatic strain in the [001] direction results in a change of the sign of the strain gradient near the top of the dot.

The biaxial strain shows more prominent effects due to the increased *In* fraction. The biaxial strain is negative within the dot and positive outside the dot penetrating deep inside the GaAs buffer from the top and the bottom dot boundaries. A strong vertical dominance of the biaxial strain field is evident in absence of SRCL, while the strain does not penetrate as deeply into the lateral directions. In the lateral direction only the wetting layer shows larger biaxial components (as expected). The introduction of the SRCL appears to spread the bi-axial strain further in the lateral direction. This is indeed evident with the line cuts of the biaxial strain through the center of the dot in the lateral direction (see Fig 3-last row). Green line shows average strain in the buffer region. Figure 3 shows a nearly constant ~5% strain throughout the SRCL layer. As the dot is surrounded by the $In_xGa_{1-x}As$ SRCL, the positive strain outside the QD decreases in the vertical direction and the negative strain inside the dot increases. The biaxial strain strongly affects the hole confinement potentials [12, 15], and therefore a reduced heavy hole (HH) and light hole (LH) splitting outside the dot and an increase (decrease) of the well depths for the HH (LH) will be observed.

The speckles in the SRCL region indicate the inhomogeneous nature of the strain in the InGaAs region which is due to the bi-modal In-As and Ga-As bond length distributions that have been observed experimentally [39] and are faithfully reproduced in NEMO 3-D [27]. While bi-axial component fringe fields are visible in the QD without a SRCL, the increasing *In* fraction in the SRCL appears to smooth out these fringe fields and homogenize the biaxial stain inside the QD.

In figure 3, we notice that the hydrostatic strain has a peak value of ~-0.094 and the biaxial strain has a peak value of ~-

0.23 for *x*=0.4 within the quantum dot region. Here we want to point out that our InAs tight binding parameters have been matched against Van de Walle [17] reference data for the hydrostatic and biaxial strains of ~-0.21 and ~-0.29 respectively. Figure 7 in reference 27 plots the InAs conduction band, heavy hole band, light hole band and split off band edges for a peak hydrostatic strain of ~-0.21 ($\varepsilon_{xx}=\varepsilon_{yy}=\varepsilon_{zz}$=-0.07) and a peak biaxial strain of ~-0.29 ($\varepsilon_{xx}=\varepsilon_{yy}$=-0.07, $\varepsilon_{zz}$=-1.084$\varepsilon_{xx}$=0.075). Our $sp^3d^5s^*$ band edges closely matched with the reference data from Van de Walle.

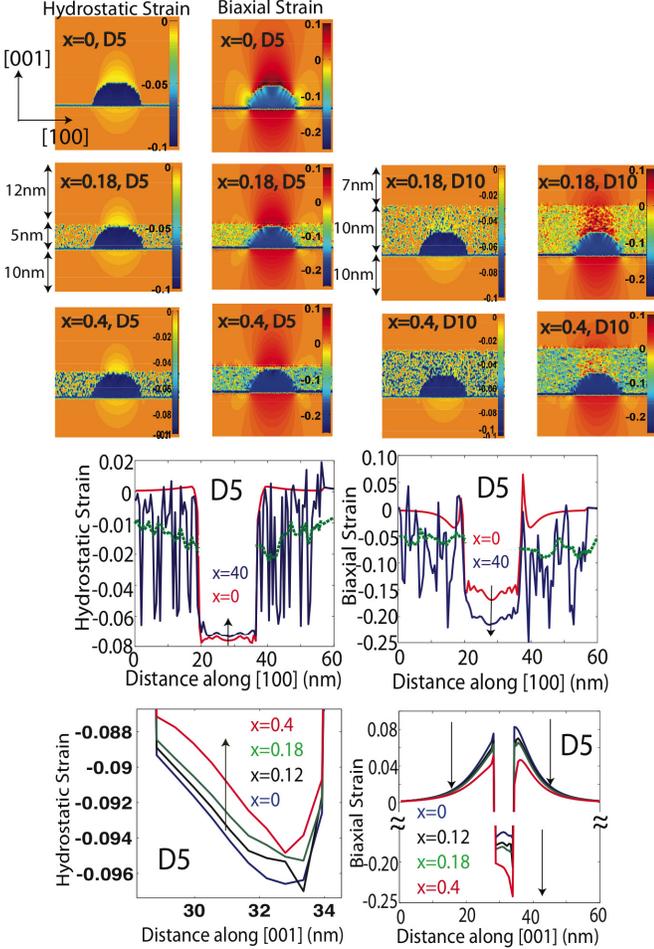

Fig. 3. Hydrostatic strain {$\varepsilon_{xx}+\varepsilon_{yy}+\varepsilon_{zz}$} (left column) and Biaxial strain {$\varepsilon_{xx}+\varepsilon_{yy}-2\varepsilon_{zz}$} (right column) for *In* fraction (x) of 0, 0.18 and 0.4 in *D5* and *D10* in the spatial region around the quantum dot. Even qualitatively the InGaAs SRCL creates a reduction in the biaxial fringing strain fields. The last two rows shows the peak hydrostatic strain (left) and the biaxial strain (right) along the [001] (upper row) and [100] (lower row) directions through the center of the InAs quantum dot as a function of distance in *D5*. The arrows are placed to guide the eyes for changes with increasing x. In [001] direction, the hydrostatic strain profile peak moves to the center of the QD making the strain profile a bit more symmetric towards the center as the *In* fraction increases. With the same *In* fraction increase a reduction of the biaxial strain around the QD and an increase inside the QD is observed in [001] direction. Line plots along [100] direction show that the SRCL spreads the hydrostatic and biaxial strains significantly in the lateral direction. Green line show the average value of strain calculated in a window of 10nm around each point. A nearly constant 5% biaxial strain is observed throughout the SRCL buffer. The biaxial strain decrease outside and increase inside points to a strong aspect ratio distortion inside the dot.

Figure 4 shows the local band edge diagrams for the lowest conduction band and the top three hole bands (heavy hole, light hole and split off) for *x*=0, 0.12, 0.18, and 0.4 in *D5* along the central line through the dot along the growth direction. Also shown are the first few confined states in the conduction and the valence bands. The optical gaps and the electronic gaps are marked on each diagram. The spatial variations in the local band edges indicate strong influence by the hydrostatic and the biaxial strain profiles.

A simplified model [14] can partially explain the strain effects. The strained electron confinement potential is determined by the hydrostatic strain (equation 1) and the strained hole confinement potentials are determined by both, the hydrostatic and the biaxial strains (equations 2 and 3):

$$\delta E_C = a_c \varepsilon_H \quad (1)$$

$$\delta E_{HH} = a_v \varepsilon_H + b_v \varepsilon_B / 2 \quad (2)$$

$$\delta E_{LH} = a_v \varepsilon_H - b_v \varepsilon_B / 2 \quad (3)$$

The hydrostatic and the biaxial deformation potential values for the InAs ($a_c$ = -5.08eV, $a_v$ = 1.0eV, and $b_v$ = -1.8eV) are well documented in the literature [17]. As the *In* fraction of the SRCL increases, the negative hydrostatic strain decreases in magnitude, reducing the positive $\delta E_C$ and shifting the conduction band down. The hydrostatic component of the strain induced band shift for the HH and LH band induces a negative $a_v\varepsilon_H$ and the reduction of $\varepsilon_H$ shifts the bands up. The biaxial strain component, however, has opposite trends for the HH and the LH band edges. The increase in the negative biaxial strain $\varepsilon_B$ inside the quantum dot creates a positive $b_v\varepsilon_B/2$ and therefore shifts the HH band up in energy and the LH band down in energy. Hence, for the HH band edge, the biaxial and the hydrostatic strain effects accumulate whereas for the LH band edge, they tend to cancel each other. For *x*=0.4, the biaxial strain effect dominates significantly over the hydrostatic strain effect and a significant change in the shape of the local LH and HH band edges can be observed.

The hydrostatic strain reduction directly shifts the electron band edge downward and it shifts HH band edges upward inside the InAs QD. The splitting between the HH and LH bands in the center of the dot increases with increasing *In* fraction from 146.9meV to 598.4meV. Since the biaxial strain has the opposite sign outside of the quantum dot and decreases with increasing *In* fraction, the trend in the HH and LH splitting has an opposite trend outside of the quantum dot and the energy splitting is seen to decrease from 221.4meV down to 145.5meV as the *In* fraction increases. Overall the LH band is shifted away from the centrally relevant energies and appears to be less important in these devices. The highest confined hole states can be expected to be primarily HH based showing strong confinement to the quantum dot region.

Figure 3 shows that the biaxial strain relaxation due to SRCL results in an increase (decrease) of the magnitude of the negative (positive) strain inside (outside) the InAs QD with increasing *x*. Since the biaxial strain is responsible for the LH and the HH band splitting [12], this increase (decrease) of the biaxial strain magnitude results in enhanced (reduced) splitting of the LH and the HH band offsets inside(outside) the QD as evident in Figure 4. From $x=0$ to $x=0.4$, a ~ 452=598.4-146.9 meV (~75.4 = 221.4 – 145.5meV) increase (decrease) in the splitting between HH and LH bands is calculated from the strain modified band edges. Similar effects have been observed in Ref 11. This increased separation of HH-LH bands causes the highest confined states to be more and more HH like states which will bind these states closer and closer to the center of the QD and lessens the penetration into the barrier.

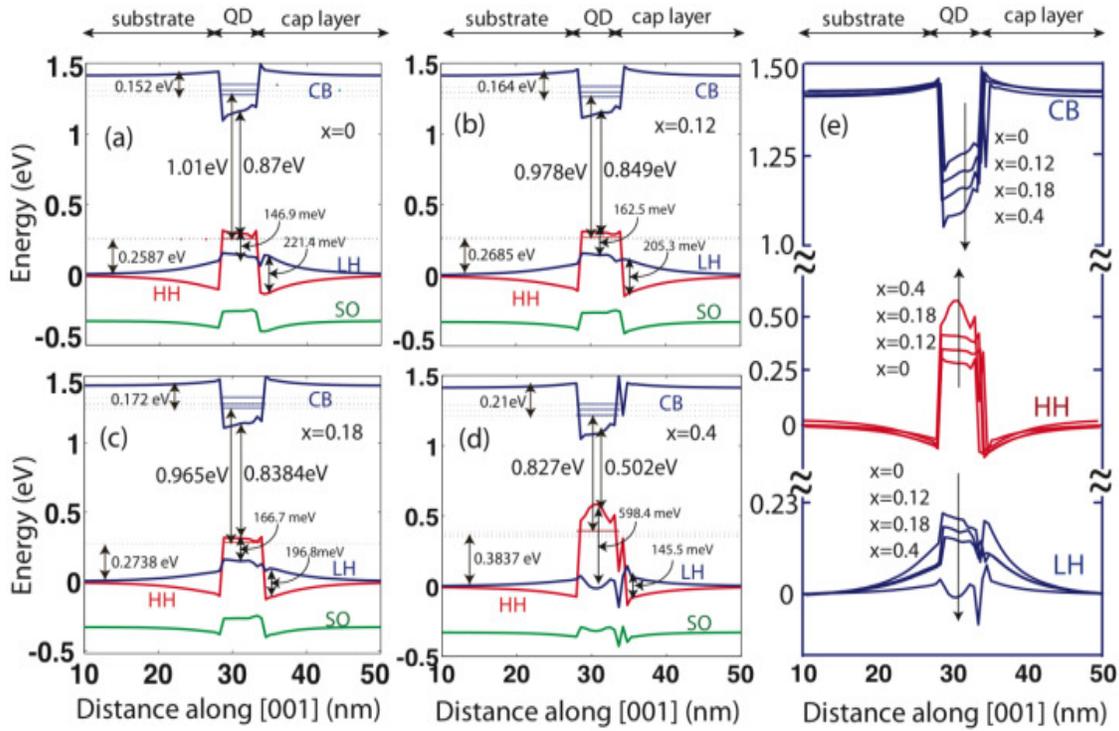

Fig. 4. (a, b, c, d) Band edge diagrams along the [001] (growth) direction through the center of InAs quantum dot for the *In* fractions ($x$) of 0, 0.12, 0.18 and 0.4 in *D5*. In each diagram, the binding energy of the first electron and the hole levels, the electronic band gap measured at the center of InAs QD, the optical band gap, and HH/LH splitting inside and outside the QD are shown. As the *In* fraction of SRCL increase, the band gaps shrink and hence a red shift of emission spectra can be predicted. (e) CB, HH and LH curves are reproduced from (a), (b), (c) and (d). Arrows are marked to indicate the direction of change in the band edges.

Figure 4 shows the binding energies for the ground electron and hole energy levels. We calculate electron binding energy of ~152meV and hole binding energy of ~258meV for $x=0$ in *D5*. Our hole binding energies are in good agreement with experimentally measured values of Itskevich et al. [21] (~250meV) and Berryman et al. [22] (~240meV) at room temperature for similar sized QDs. As an additional verification for our electron binding energy, we point to another experiment. Tang et al. [20] reported an electron binding of ~80meV for a QD with an estimated base diameter of 13 to 17nm. For a base diameter of 15nm and height of 3.5 nm, we calculate an electron binding energy of ~91meV. This agreement with experimental electron and hole binding energies quantitatively validates the NEMO 3D simulation capabilities.

A similar QD system was simulated in Ref. 19 using an atomistic pseudo-potential method. Using a QD base diameter of 25.2nm and QD height of 3.5nm, the authors reported the electron and hole binding energies as 271meV and 193meV, respectively. We simulate the same sized QD and found that the electron and hole binding energies are ~181meV and ~289meV respectively. The pseudo-potential calculations, for the same sized QD, result in electron confinement energies that are too low and hole confinement energies that are too high while our tight-binding NEMO 3D results are much closer to the experimental values of Berryman et al. [22] and Tang et al. [20]. We attribute the NEMO 3-D agreement with technologically relevant room temperature data to the intensive tuning efforts [10, 27, 28, 29] of the InAs and GaAs constituent materials to available experimental room temperature bulk data. Another critical advantage of a local basis representation used in NEMO 3-D over a plane wave basis used in pseudo-potentials is that the representation of arbitrarily disordered systems does not require an expert user to tune the number of included plane waves in the system. The computational burden is determined in NEMO 3-D purely by the size of the system, not the degree of disorder. This has particular relevance for the creation of a nano-TCAD tool to be used by non-theory experts as envisioned for a community on nanoHUB.org.

As a general trend we observe here that the introduction of the SRCL increases the binding energies with the *In* fraction ($x$). A 58meV increase from 152meV to 210meV and a ~125meV increase from ~258meV to ~383meV in the electron and the hole binding energies, respectively, is calculated from $x=0$ to $x=0.4$.

Figure 3 shows for $x=0.12$, 0.18, 0.28, and 0.4 an average hydrostatic strain magnitude reduction of 0.00197, 0.003, 0.0033, and 0.00494 and an average increase in the magnitude of biaxial strain within the QD is 0.01436, 0.018, 0.0508, and 0.064, respectively. Corresponding to these hydrostatic and biaxial strain changes, Eq. (1) gives an average change in the strain-modified conduction band edge of 10meV, 15.25meV, 16.9meV, and 25.114meV and Eq. (2) for the HH band edge, a corresponding change of 14.9meV, 20meV, 49.09meV, and 63.29meV. These changes in the confinement potentials are expected to translate directly into the corresponding changes of the electron and the hole ground state energy levels and hence the optical band gap. However, the simple model predicts, at $x$=0.4, a ~88meV (25.114+63.29 meV) strain induced red shift of the emission spectra while a ~183meV change of the optical gap is simulated with NEMO 3-D simulation $x$=0 and $x$=0.4 in figure 4. We therefore conclude that at large values of the $x$, the strain contributes less than half of the overall red shift of the emission spectra. Another physical effect is needed to explain the remaining shift in the emission spectra. In the next section, we will explain the effect of aspect ratio (AR) variation in the overall red shift of emission spectra.

### A. Aspect ratio change of the quantum dot

The argument that the hydrostatic and biaxial strain components shift the conduction band and HH band edges, while applicable to bulk semiconductors, is not enough to account for the total shift in the optical absorption energy. We show here that the optical absorption energy is not only influenced by the bottom of the conduction band and the top of the valence band but also by the physical shape of the quantization potential.

In the absence of the SRCL, the in-plane strain ($\varepsilon_{xx}$, $\varepsilon_{yy}$) is compressive and the vertical strain ($\varepsilon_{zz}$) is tensile. The reason for the tensile vertical strain ($\varepsilon_{zz}$) is that the quantum dot is flat and the larger in-plane compression results in vertical expansion due to the Poisson effect. Self-assembled InAs QDs form on GaAs to minimize the bi-axial strain in the growth layer. Only a thin strongly biaxially strained wetting layer remains. Capping the InAs wetting layer and QD with the InGaAs alloy introduces more In-As bonds into the system which push against the lattice constant imposed by the GaAs substrate. The whole InGaAs SRCL is biaxially compressively strained to accommodate the GaAs substrate. That implies that the InAs QD can no longer compensate by expanding against the GaAs cap, with a smaller lattice constant. The InAs QD must compensate by expansion into the growth direction. Due to these effects, biaxial strain ($\varepsilon_B$) inside the quantum dot is increased and hydrostatic strain ($\varepsilon_H$) is decreased for increasing $x$ in the SRCL as shown in figure 3. In conclusion, with the introduction of the SRCL, the InAs atoms inside the quantum dot align themselves more closely with the vertical atom planes in the underlying substrate, push less laterally against the surrounding buffer, and compensate the loss of volume by vertical expansion. The quantum dot height increases and the quantum dot width decreases slightly as shown in the inset of Figure 5 resulting in a change in aspect ratio (height/base) as indicated in Figure 5. The quantum dot changes shape! Appendix A derives [16] from a very simple particle in a box problem that for low aspect ratio QDs (aspect ratio < 0.3), small increases in QD height decrease the band gap more than a corresponding increase in base length. The following paragraphs elaborate on the effect of the change in the quantum dot shape and relate it quantitatively to the observed change in optical band gap.

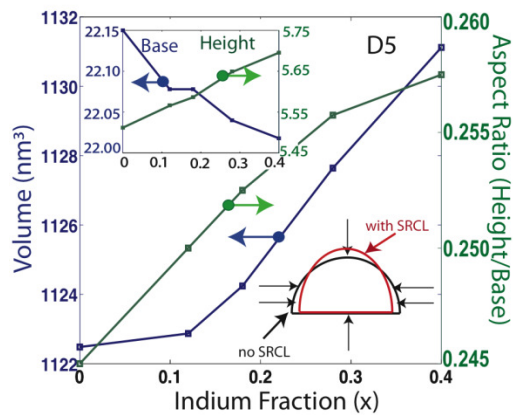

Fig. 5. Volume of the InAs QD and the aspect ratio (height/base) are shown as a function of the *In* fraction (*x*) in the SRCL in *D5*. As the *In* fraction increases, the dot height and the aspect ratio increases. Inset shows the base and the height of the InAs QD as a function of the *In* fraction in the SRCL. As the *In* fraction increases, the height increases and the base length decreases. Decrease in base length results in small blue shift and increase in height of QD results in large red shift of emission spectra.

In the presence of the SRCL, the reduction of the base diameter of the InAs QD increases the optical band gap of the system (a blue shift of the emission spectra). On the other hand, the increase in the height of the InAs QD has the opposite trend: it tends to decrease the optical band gap of the system (the red shift of the emission spectra). The QDs studied in this paper are the low aspect ratio lens shaped QDs and are much more sensitive [16] to the height changes as compared to the base diameter changes. The increase in the optical band gap due to the changes in the base diameter are therefore negligible compared to the decrease in the optical band gap due to the increase in the height of the QD. As a result, the size changes of the QD in the presence of the SRCL overall contribute in the red shift of the emission spectra. For the large values of the *In* fraction (*x*), this effect becomes a significant source of the energy shift. From *x*=0 to *x*=0.4, It is calculated from the NEMO 3-D simulations (Appendix B) that the AR change causes a red energy shift of ~ 90meV. This is a significant (~50%) contribution in the total optical band gap shift of ~183meV between *x*=0 and *x*=0.4.

Our numerical results can be verified through a simple single band effective mass particle in a box model (ignoring the strain effects). It predicts a decrease of ~60meV in the optical band gap for the base and height changes in Figure 5 (see Appendix A). Figure 6(b) compares the NEMO 3-D results with the results from simple analytical models. The strain contribution is calculated from equations (1) and (2) and the data from Figure 6(a). AR contribution is calculated from data of Figure 5 and the procedure of Appendix B. The AR contribution shows a strong non-linear behavior while the strain shows a roughly linear behavior. The total change in the optical emission wavelength calculated from the analytical models shows a non-linear behavior in a reasonable qualitative agreement with NEMO 3-D results. The simplified model in Appendix A indicate that the aspect ratio change of the quantum dot indeed results in a significant change in quantum confinement and that in turn changes the band gap significantly. Clearly this change in aspect ratio is strain induced, but change in the quantum dot shape is a bit of a surprising effect beyond the typical band gap change due to strain.

Figure 6 (c) plots the first five electron and (d) the first three hole energy levels as a function of the SRCL $In$ fraction ($x$) for $D5$ and $D10$. Three critical physical issues can be extracted from figure 6: (1) Dependence of SRCL thickness (2) Applicability of lasing devices (3) Relation of nonlinear biaxial strain and hole energies, as discussed in the three subsequent paragraphs.

Increasing the SRCL thickness $D$ from 5nm to 10nm does not change the optical gap significantly. At $x=0.4$, $e1(D5)$ – $e1(D10)$ is only 7.5 meV and $h1(D5)$ – $h1(D10)$ = 14.3 meV. This is because once the QD is covered with the $In_xGa_{1-x}As$ SRCL, further increase in the thickness $D$ does not change the strain relaxation or the QD size much. This effect is also presented in Ref 14.

The energy difference $e2$-$e1$ is quite important for laser applications. This difference should not decrease with increasing SRCL effects to avoid undesirable occupation of higher excited states [5]. Figure 6(c) shows that this difference (~ 36 meV) is almost independent of $x$. Hence, our atomistic tight binding model predicts that the change of the $In$ fraction of the SRCL does not limit the possible use of this device for the long wavelength laser applications.

The change in electron energy levels is approximately linear with increasing $x$ corresponding to the linear behavior of the hydrostatic strain in figure 6(a) and equation (1). The hole energy levels show an abrupt jump for $x$ above 0.18 correlated with the nonlinear behavior of the biaxial strain in figure 6(a). Since the hydrostatic strain impact for both electron and hole energy levels is the same (equations 1, 2 and 3), this additional change above $x=0.18$ can be explained in terms of the biaxial strain component as follows: Figure 6 (a) plots the hydrostatic and biaxial strain as a function of $In$ fraction $x$ at the center of InAs QD. The hydrostatic strain decrease is almost linear with $x$. For the $In$ fractions ($x$) above 18%, a large increase in the negative biaxial strain within the InAs QD contributes in the significant change in the hole confinement potentials and energy levels.

The origin of the nonlinear behavior of biaxial strain in Figure 6(a) can be explained in terms of nonlinear change of the bond lengths under strained environment. Figure 7(a) plots the mean bond lengths of $In_xGa_{1-x}As$ alloy as a function of $In$ composition ($x$). The blue lines are experimentally reported data from Mikkelsen and Boyce [39] for a bulk $In_xGa_{1-x}As$ alloy. The black lines are from NEMO 3-D simulations for a bulk $In_xGa_{1-x}As$ alloy [27]. However, when an alloy is placed under strain distortion, the linear nature of bi-modal bond length changes. Red lines are from NEMO 3-D data for a strained $In_xGa_{1-x}As$ 5nm thick quantum well placed on and capped by GaAs. The bond length distribution is still bi-modal, however the In-As and Ga-As curves approximately parallel to each other. The strain applied by the GaAs substrate changes the linear bond length distortion observed in bulk to a behavior virtually independent of $In$ fraction. Similar results were reported in Ref. 38.

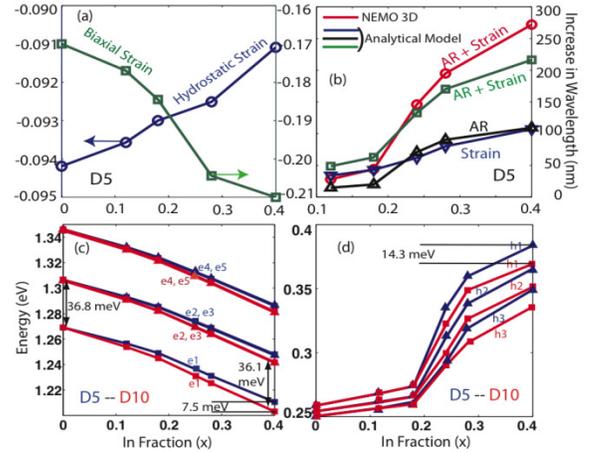

Fig. 6. (a) Hydrostatic and biaxial strain as a function of $In$ fraction of SRCL at the center of InAs QD. As $x$ increases, the decrease in the hydrostatic strain is almost linear with $x$. A significant increase in the biaxial strain can be noted for the values of $x$ above 0.18. (b) Results of NEMO-3D are compared with simple analytical model results. Red line is from NEMO-3D simulations and green, black and blue lines are from simple analytical expressions. Strain contribution is computed from equations (1), (2) and (3) using strain values of figure (a). The AR contribution is calculated from the procedure of appendix B. NEMO-3D results qualitatively match with the results from simple analytical model; First five electron energy levels (c) and first three hole energy levels (d) as a function of the $In$ fraction ($x$) of the SRCL in $D5$ (blue) and $D10$ (red). Experimental[3] values of In fraction (0, 0.12, 0.18, 0.24, 0.28, 0.4, and 0.45) are used. As $x$ increases, the electron energy level decreases and hole energy level increases and the optical band gap shrinks. A total change of ~200meV (in $D5$) and ~198meV (in $D10$) is calculated between $x=0$ and $x=0.45$.

An analytical model was developed in Ref. 37 for strained alloys. We compare our NEMO 3-D results for strained QW (blue line in figure 7(b)) with the analytical model [37] (red line in figure 7(b)). A close qualitative match is found, to within 0.73%, validating the NEMO 3-D model. A critical issue that cannot be resolved with an analytical model is the influence of an embedded quantum dot on the bond distribution. The black line plots the average crystal bond length within the QD region for a QD placed inside the $In_xGa_{1-x}As$ alloy ($D5$). The bond lengths increased slightly but have the similar nonlinear trend.

Figure 7(c) plots the average bond length as a function of distance along [001] direction through the center of quantum dot in $D5$. Interestingly, the bimodal distribution has a spatial dependence within the quantum dot. Also we can see the non-linear change of bond lengths within the QD as $x$ increases from 0 to 0.4. Hence we conclude that in strained alloys, the linear interpolation of bond length becomes invalid, and an atomistic study is necessary to capture the nonlinear behavior of bond lengths.

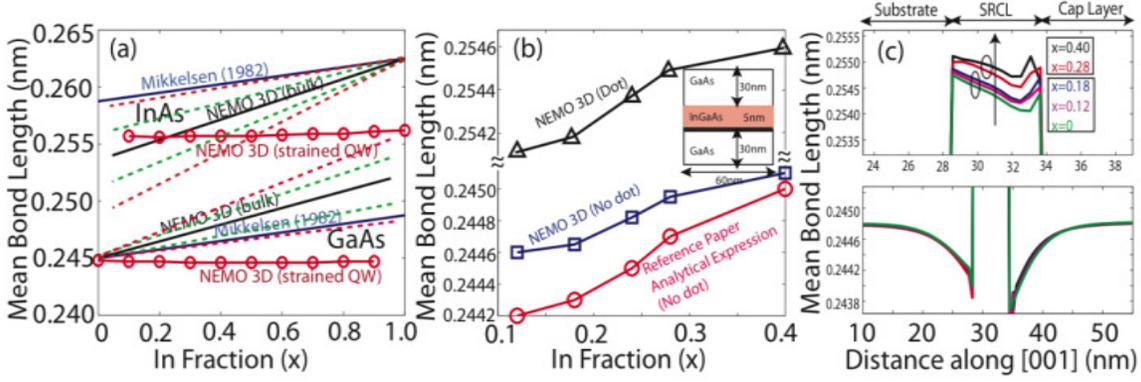

Fig. 7. (a) In-As and Ga-As bond length average as a function of In Fraction ($x$). Blue lines correspond to experimental data reported by Mikkelsen and Boyce (1982). Black lines are for Bulk $In_xGa_{1-x}As$ material reproduced from Reference 27. The green and red dotted lines show one and two ± standard deviations respectively around the data points. The NEMO 3-D data matches the Mikkelson's experimental data closely within two standard deviations. Solid red lines are for strained QW ($In_xGa_{1-x}As$ QW inside GaAs buffer) using NEMO-3D. (b) Mean crystal bond length as a function of In Fraction $x$ of SRCL. Red line is the calculated bond lengths from analytical expressions of reference 37 for 1D QW. Blue line is for NEMO 3D simulations for 1D QW. Black lines are for NEMO 3D simulations including QD inside QW. (c) Average bond lengths as function of distance along [001] in $D5$. A nonlinear change in bond length inside QD can be observed as In Fraction is changed from 0 to 0.4.

Figure 8 (a) shows the calculated optical band gap and the emission wavelength as a function of SRCL In fraction ($x$) in $D5$. The black curves are shown from experiments [3]. As the value of $x$ increases, the optical band gap decreases and the emission wavelength increases. A 1.53μm wavelength is calculated for $x=0.45$ which closely matches the experimental value of 1.52μm.

An astonishing agreement with experiment is observed. We emphasize here again that there were no adjustment to any material parameters [10] performed in NEMO 3D to match the experiment. In particular we note that the simulation quite nicely models the non-linear behavior of the experimental data.

The sections above describe the contribution of the strain and AR variation in the overall red shift of the emission spectra using a simulator based on atomistic tight binding parameters [10]. The strain and AR variations each contribute about half of the energy shift. To validate our results with a simple analytical particle in a box type single band effective mass relation, we performed calculations (see Appendix C) using data from figure 3 and 5. The summary of the results is shown in Table 1. Our simulation results have trends similar to those predicted by the simple effective mass model, which allows us to understand the results more intuitively.

Table 1: Comparison of the strain and AR variation contributions in the overall red shift of the emission spectra:

### A. Atomistic interface effect

Experimentally, it is very hard to find the exact composition of QDs. Particularly, the abrupt interface between InAs QDs and the surrounding GaAs buffer, assumed in most of the literature may not be true. To investigate the effect of QD interface softness on the emission spectrum, three different samples of QD were simulated. The QD consisted of an inner dome of diameter 18nm composed of pure $In_{x1}Ga_{1-x1}As$ ($x1=1$, InAs) material. The dome was then covered by 1nm thick layers of each $In_{x2}Ga_{1-x2}As$ and $In_{x3}Ga_{1-x3}As$ (see inset in Figure 7(b)). In the first sample $D5^{(1)}$, $x1$, $x2$ and $x3$ all have value of 1. This corresponds to a pure InAs QD which was previously labeled as $D5$. In the second sample $D5^{(2)}$, $x1$, $x2$, and $x3$ have values of 1, 0.9, and 0.8 respectively. In the third sample $D5^{(3)}$, $x1$, $x2$, and $x3$ have values of 1, 0.7 and 0.6 respectively. The emission wave lengths for all three compositions of QD are plotted in figure 8(b). The grading of the QD/buffer interface produces only a small blue shift in the emission spectrum. The larger the interface gradient, the larger the blue shift. A maximum blue shift of 47nm ($D5^{(2)}$ at $x=0.45$) and 70nm ($D5^{(3)}$ at $x=0.45$) is observed. The results give a quantitative measure of the interface effects on the emission spectra and indicate that the interface details are not of utmost criticality in this particular system.

### B. Quantum dot size variation effect

Experimentally, it is very hard to create a quantum dot of precise size or to measure the exact QD size after capping. Modeling such devices should therefore also consider the effects of size fluctuations. To quantify the effect of the QD size on the wave length of emission spectra, we simulated two different size variations of the $D5$ (Base diameter 20nm, Height 5nm). Figure 8(c) shows the emission wave length as a function of In fraction ($x$) of the SRCL for (1) Base Diameter=21nm and Height=5nm and (2) Base Diameter = 20nm and Height = 5.5nm against $D5$. In both cases, the emission spectra wave lengths shifts towards higher values (red shift). An average red shift of ~40nm and ~120nm is calculated for case (1) and (2) respectively. Figure 8(c) compares our results with the experimental data [3] (black lines). The three black lines are for the same InGaAs compositions grown by low-pressure MOCVD using trimethylindium (TMI), trimethylgallium (TMG), triethylgallium (TEG), and tertiarybutylarsine (TBA) as the source materials at the total pressure of 76 Torr. Three samples of the same QDs were grown by changing the growth conditions of InAs QDs, and the source materials of the GaAs capping layer. The conditions of the TBA partial pressure, the growth rate of InAs QDs, and the source materials of GaAs capping layer are shown in Table 1 of Reference 3. The comparison of our results with these experimentally reported curves (Figure 4 in reference 2) show that a QD with base diameter 21nm and height 5nm matches the experimental curve more closely. Hence, we conclude that the QD in the experiment might have a height of 5nm and base diameter of around 21nm.

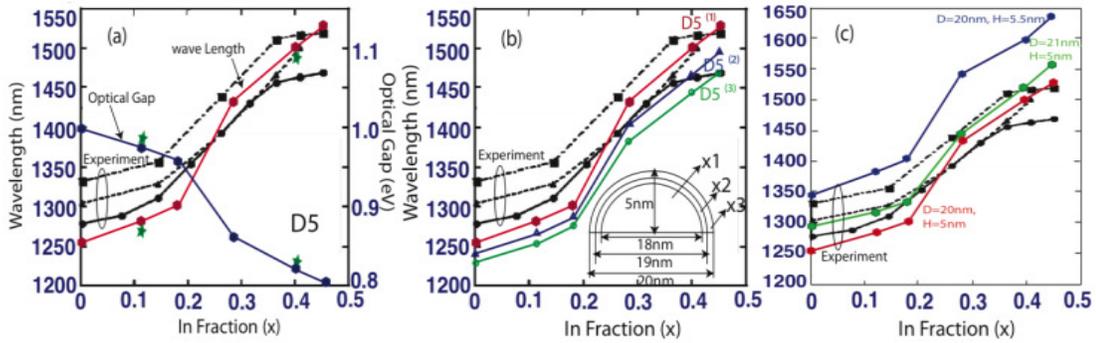

Fig. 8. (a) Optical band gap and the emission wavelength as a function of the *In* fraction (*x*) in the SRCL in *D5*. The black curves are experimental values from Ref 3. We calculate an emission wavelength of 1.53um for experimental value of 1.52um at *x*=0.45. At *x*=0.12 and 0.4, the optical gap and the emission wavelength are marked (green star) for a $In_xAl_{1-x}As$ SRCL.(b) Emission wave length as a function of *In* fractions (*x*) of the SRCL. Three different quantum dot compositions are considered. $D5^{(1)}$ is pure InAs QD with x1=1, x2=1 and x3=1. In $D5^{(2)}$, QD has mixed composition with x1=1, x2=0.9 and x3=0.8. In $D5^{(3)}$, the QD has a mixed composition with x1=1, x2=0.7 and x3=0.6. The inset shows the structure of the QD. Black curves are reproduced from the experiment[3] for the reference. (c) Emission wave length as a function of *In* fractions (*x*) of the SRCL. Two size variations of the QDs are considered: (1) Base diameter is changed to 21nm and height is fixed to 5nm (2) Height is changed to 5.5nm and base diameter is kept constant to 20nm. The height and base diameter increase causes a red shift in the emission spectra.

*A. Change in electronic confinement*

The sections above showed that the SRCL strongly modifies the strain profiles. Stepping back one notices that a material is introduced that effectively reduces the electronic confinement, since the InGaAs band gap is smaller than the GaAs confinement buffer band gap. One could argue [5] that this reduction in confinement is the reason for a reduced band gap, since the wave functions can leak out from QD. Here we test this argument by inserting a strain reducing buffer that has virtually the same strain effects as InGaAs, but a larger gap. We introduce an $In_xAl_{1-x}As$ SRCL for *x*=0.12 and *x*=0.4 in *D5*. Everything else is kept unchanged. The change in the optical gap is found to be only 14meV (for *x*=0.12) and 13.4meV (for *x*=0.4) shown as green stars in Figure 8(a). Since the $In_xAl_{1-x}As$ band gap is larger (by 1.3eV) as compared to the $In_xGa_{1-x}As$, a small change in the optical gap indicates that the reduced barrier height on the top side of the QD has only a minor contribution in the significant observed red shift. This is because the electron and the hole states are strongly localized within the band edge diagram wells (see Figure 4 and, for 3-D volume representations, Figure 9). In fact, especially for the holes, it can be observed that the strain confines the wave-function more tightly into the quantum well. This effect is also observed in experiment [7]. In contrast, Ref 5 uses a single band effective mass model and predicts that the reduced barrier height on the SRCL side has a significant contribution in the overall red shift of the emission spectra. Our result clarifies the physical picture and show that the reduced confinement has only a negligible contribution in the observed red shift of emission spectra. Here we also conclude that an atomistic model applied to realistic structures rather than a continuum effective mass k•p model is critical in the understanding of quantum effect devices at the nanometer scale.

*B. Wave function confinement and symmetries*

Figure 9 plots the calculated first four electron (*e1, e2, e3, e4*) and the first two hole (*h1, h2*) wave-functions squared for *x*=0, 0.18, and 0.4 in *D5* and *x*=0.4 in *D10*. The ground hole state *h1* is found to be more localized at the center of the InAs QD with increasing values of *x* (~83% for *x*=0 and ~97% for *x*=0.4 in *D5*). The increase in the *In* fraction has a strong effect on the localization of the first excited hole state which moves from the QD perimeter to the QD center. The electron ground states do not bear much change in terms of localization (~79.7% for *x*=0 and ~80% for *x*=0.4 in the *D5*).

The increase of a In=0.4 SRCL height from 5 to 10nm does not affect the symmetries of the state shown in the bottom two rows of Figure 9 at all. This is not very surprising as the Eigen energies of these states vary only slightly as shown in Figure 6 (c), (d).

A slight clockwise rotation is found for the excited states *e2* and *e3* with an increased *In*-fraction. This rotation can be explained in terms of the hydrostatic strain relaxation at the top of the InAs QD. The negative gradient in the hydrostatic strain causes an unequal stress in the zincblende lattice structure along the depth breaking the equivalence of the [110] and the [1$\bar{1}$0] directions. This breaks the degeneracy of the first excited state called a p-state aligning the *e2* state in the [1$\bar{1}$0] direction and the *e3* state in the [110] direction [18, 23]. The reduction of the hydrostatic strain in the growth direction and the shift of the peak of the hydrostatic strain from the top of the dot towards the center of the dot as evident from figure 3 causes this slight rotation in the wave-functions of the *e2* and the *e3* states.

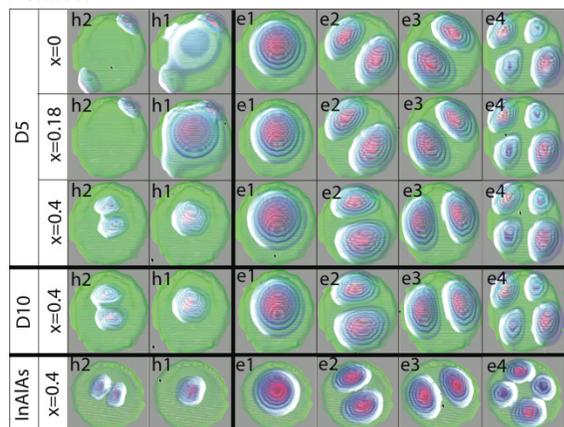

Fig. 9. Top view of the first four electron (e1 - e4) and first two hole (h1-h2) wave-function magnitudes for x= 0, 0.18 and 0.4 in *D5*, x=0.4 in *D10*, and 5nm thick $In_{0.4}Al_{0.6}As$ SRCL (last row). The green region shows the boundaries of the InAs QD. The blue and the red color show the intensity of magnitude (blue is lowest and red is highest).

*C. Neglected physical effects*

Although the tight binding approach and NEMO 3-D can indeed treat excitonic effects [31] and piezo-electric effects [9, 18] we neglect these effects in this work. Excitonic effects lower the band gap by a typical value of 10-30meV corresponding to a change of ~15-47μm, which is negligible in comparison to the overall experimental uncertainty as demonstrated by the fluctuations due to interface inter-diffusion / softness and quantum dot size uncertainty. The piezo-electric effects have importance in the ordering and energy splitting of the excited states and are of order 5meV [18]. Since we are not critically interested here in the polarization of the emitted or absorbed light from excited states, we also neglect this effect.

*D. Nano-TCAD deployment on nanoHUB.org*

NEMO 3-D was originally developed at the NASA Jet Propulsion Laboratory from 1998-2003. It continues to be developed at Purdue University with algorithmic and physics refinements. In September 2005 an educational version of NEMO 3-D was released on nanoHUB.org which allows the computation of particle in a box states for various geometries, their intra-band transitions, and the visualization of 3-D volume rendered wave-functions (see sample screen shots in Fig. 10). The computation happens extremely rapidly since the educational version only contains the single "s" orbital model, which corresponds to an effective mass model. The growth and use of the tool on nanoHUB.org has been strong (Fig. 10 inset). Over 1,660 users have run over 15,500 simulations since the release of the "Quantum Dot Lab". Stabilization of grid computing resources now allows us to deploy the full NEMO 3-D version which requires about 10 hours on 60 CPUs for the problem sizes presented here.

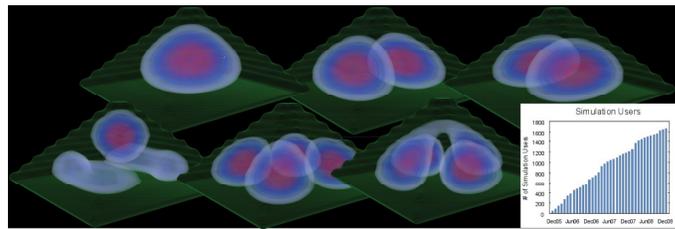

Fig. 10. First 6 eigen states of a pyramidal quantum dot computed in the "Quantum Dot Lab" on nanoHUB.org. The pyramidal symmetry breaks the expected 4th $p_z$ state into one with a topmost lobe in the apex and two side lobes. The 6th state shows interesting arching of the wave-function lobes. Users can perform 3-D volume rendered visualization interactively themselves. Also shown on the bottom right is the growth of users of the "quantum Dot Lab" to over 1,660 since its release in November 2005 for the Supercomputing Conference, where it was used for the first time as an outreach activity for high school and college teachers.

IV – CONCLUSIONS

A detailed qualitative and quantitative analysis of the strain relaxation, the QD volume, the aspect ratio changes, and the reduction of the barrier height on the SCRL side was presented. The hydrostatic and the biaxial strain relaxations in the growth direction change the electron and hole confinement potentials and contribute to the red shift of the emission spectra. The volume and the aspect ratio of the QD increase as the indium concentration of the SRCL increases. For large values of the *In* fraction in $In_xGa_{1-x}As$ (above 0.2), the increased aspect ratio becomes a significant source of the energy shift in the strain-reduced QDs. The SRCL induced barrier height reduction, buffering the InAs QD, has only a negligible effect in the overall energy shift, as shown by an InAlAs comparison. The quantum dot boundary and the surrounding buffer interface softness cause a small blue shift in the emission spectra. A brief size variation study points to a slightly larger base width than reported experimentally. Both the interface softness study and the size variation study show that the physical processes presented here are robust against experimentally unavoidable imperfections. The NEMO 3D tight binding model achieves an astonishing agreement with experimental data reproducing the nonlinear change in the optical band gap with the introduction of a SRCL. No parameter [10] adjustment was done for this multi-million atom electronic calculation. The close quantitative agreement of our theoretical analysis with experimental evidence [3] enables us to gain significant insight into the physics of this system. Continuum strain and electronic structure theories, such as an effective mass model or a k•p method, that do not account for the atomistic details, cannot achieve such detailed explanation or might even lead to incorrect conclusions.

ACKNOWLEDGEMENTS

Authors are thankful to Prof. Timothy Boykin (University of Alabama, Huntsville USA), Prof. Lloyd Hollenberg (University of Melbourne, Australia), Prof. Mincheol Shin (Information and Communications University, Daejeon Rep. of Korea), and Dr. Benjamin Haley (Purdue University, West

Lafayette IN) for reviewing the manuscript and providing valuable suggestions. This work was carried out in part at the Jet Propulsion Laboratory (JPL), California Institute of Technology, under a contract with the National Aeronautics and Space Administration (NASA). Muhammad Usman is funded through Fulbright USAID (Grant ID # 15054783). nanoHUB.org [24] computational resources operated by the Network for Computational Nanotechnology (NCN) funded by the National Science Foundation were used in this work.

## APPENDIX – A

The effects of a change in the quantum dot aspect ratio can be estimated through a very simple effective mass model. Starting from a simple particle in a box model:

$$E_{nx, ny, nz} = \frac{h^2}{8m^*}\left(\frac{n_x^2}{D^2} + \frac{n_y^2}{D^2} + \frac{n_z^2}{H^2}\right) \quad (1)$$

(Here D is diameter and H is height of the box)

For the ground state energy, $n_x = n_y = n_z = 1$,

$$E_0 = \frac{h^2}{8m^*}\left(\frac{2}{D^2} + \frac{1}{H^2}\right) \quad (2)$$

Now the relative variation of the ground state energy as a function of D and H is:

$$\frac{\Delta E_0}{\Delta H} = \frac{\Delta E_0}{\Delta D}(0.5 AR^{-3}) \quad (AR = H/D) \quad (3)$$

Hence for the low aspect ratio QDs (AR < 0.3), the variation of ground state energy is much more sensitive to height than the diameter [16].

The quantitative effect of the aspect ratio change depicted in Figure 5, can be estimated with this simple model as follows:
m* for electron = $0.023 m_o$, m* for holes = $0.41 m_o$, $m_o$ = 9.1e-31 kg (masses taken from Ref. 26)
h = 6.626e-34 joules/sec, $E_{gap}$ of InAs = 0.354 eV (neglecting strain effects)

Now using equation (1):

For D = 22.15nm and H = 5.43nm
Optical Band Gap = $E_0$ (electron) + $E_0$ (hole) + 0.354 = ~1.01 eV

For D = 22.026nm and H = 5.69nm
Optical Band Gap = $E_0$ (electron) + $E_0$ (hole) + 0.354 = ~0.95 eV

Hence,
Change in Optical Gap = 1.01 – 0.95 = ~60 meV
Hence, with the simple effective mass model and the changes in quantum dot shape from figure 5, the effect of QD shape change is estimated to 60meV. That is comparable to the analytical estimate of 88meV band gap reduction due to strain alone. This simple effective mass model does not appropriately account for the effects of non-parabolicity. Also the effects of strain deformation are not included. Appendix C presents a simple numerical model that includes these effects.

## APPENDIX – B

To calculate the contribution of AR variations alone, the electronic structure of a free standing dome and box-shaped InAs QD (without any surrounding GaAs buffer) is computed using NEMO 3-D. The strain calculations are turned off to exclude the strain contribution. The optical band gap is computed from the ground electron and hole energy levels. The following cases of InAs QD dimensions, taken from figure 5 (inset), are simulated separately:

Case 1: diameter=22.018nm, height=5.5nm ➔ The optical band gap is 1.03eV for the dome and 0.43eV for the box shape
Case 2: diameter=22.150nm, height=5.5nm ➔ The optical band gap is 1.0399eV for the dome and 0.431eV for the box shape
Case 3: diameter=22.018nm, height=5.69nm ➔ The optical band gap is 0.931eV for the dome and 0.397eV for the box shape

Hence we find that the reduction in the base diameter results in approximately blue shifts of 1.0399 - 1.03 = ~10 meV for the dome shape and ~1meV for the box shape and the increase in the height results in about 1.03 – 0.931 = ~100meV and ~33meV red shift of the emission spectra, respectively. Compared to the simple effective mass model in Appendix A, we include the effects of non parabolacity, band coupling accounts for about 27meV blue shift and QD dome shape about 67meV red shift.

## APPENDIX – C

Appendix A estimates the effect of aspect ratio from a simple effective mass model with parabolic band assumption. InAs is known to be a very non-parabolic material. In this Appendix, we estimate the effects of non-parabolicity of InAs bands, strain, and aspect ratio through the following simple numerical experiment:
1. From a single bulk InAs unit cell (with and without strain), we compute the bulk E(*k*) dispersion.
2. Now assuming a rectangular, infinite wall quantum dot, we quantize the momentum vector *k* in *x*, *y*, and *z* directions.
3. Extract the band gap and the local effective masses in the vicinity of the quantized ($k_x$, $k_y$, $k_z$) tuple on E(*k*) diagram.
4. By varying ($k_x$, $k_y$, $k_z$) according to AR variations, we can estimate its effect.
5. From the strain distortions of the InAs cell, we can explore the effect of strain.

The procedure is as described below:

The basic *E-k* dispersion relation is:

$$E_{k_x, k_y, k_z} = \frac{h^2}{8\pi^2}\left(\frac{k_x^2}{m_x^*} + \frac{k_y^2}{m_y^*} + \frac{k_z^2}{m_z^*}\right) \quad (4)$$

where $k_x = \frac{\pi}{L_x}$, $k_y = \frac{\pi}{L_y}$, $k_z = \frac{\pi}{L_z}$ and $L_x$, $L_y$, and $L_z$ are dimensions in *x*, *y*, and *z* directions, respectively.

Correct effective masses in *x*, *y*, and *z* directions at the point of dispersion ($k_x$, $k_y$, $k_z$) are calculated from the band structures of InAs in *x*, *y*, and *z* directions respectively. The point of dispersion ($k_x$, $k_y$, $k_z$) is computed from dimensions of InAs quantum dot taken from figure 5. Figure 11 shows two sample cases. The correct effective masses and corresponding optical gaps computed from equation 4 are given in table 2.

Using correct effective masses from table 2, we conclude that:
AR contribution = 0.56 – 0.482 = ~ 78meV
AR + Strain contribution = 0.831 – 0.687 = ~144meV
Strain Contribution = 144 – 78 = ~66meV
We also computed the band gaps directly from E(*k*) diagram instead of using equation 4. This is done by plotting E(*k*) diagram in [111] direction and looking at the point of dispersion ($k_x$, $k_y$, $k_z$). These numerically computed band gap variations are:
AR contribution = ~72meV
AR + Strain contribution = ~132meV
Strain Contribution = 132 – 72 = ~60meV

These results from simple analytical expressions with correct effective masses show a close qualitative match with our simulation result and provide a qualitative validation of our simulator. Also ~12meV difference between AR contribution calculated in Appendix A and C highlights the importance of using correct effective masses taking into account the non-parabolicity of the bands.

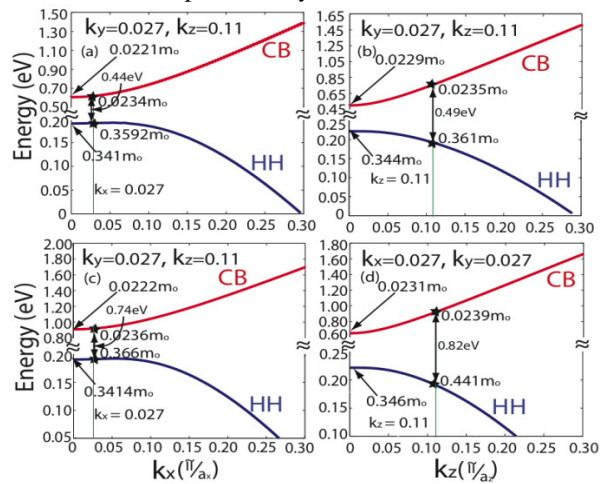

Fig. 11. Calculation of the effective masses in $k_x$, $k_y$, and $k_z$ direction from the bulk band structure of InAs. Only one case for $k_x=k_y=0.027$ and $k_z=0.11$ is shown without strain (a), (b) and with strain (c), (d). The nonparabolicity of the bands is quite evident from the diagrams. Effective masses at the dispersion point and origin are also shown. Strain increases the band gap as well as the effective masses.

Table 2: Effective masses computed from InAs band structure are shown for various cases studied:

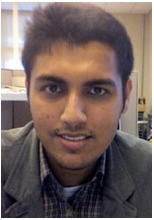

**Muhammad Usman** received his BSc. (honors) And MSc. Degrees in Electrical Engineering from University of Engineering & Technology, Lahore Pakistan with distinction in 2003 and 2005 respectively. He joined Purdue University in 2005 as a PhD student with Fulbright fellowship. Currently, he is working in Prof. Klimeck group as a PhD student. His area of research is theoretical modeling of quantum dots for optical and quantum computing applications. He is a student member of IEEE, APS and MRS societies.

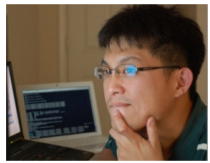

**Hoon Ryu** received the B.S. degree from Seoul National University, Seoul, South Korea in 2003, and M.S. degree from Stanford University, Stanford, CA, in 2005, respectively. He was with the Semiconductor Division, Samsung Electronics, as a Research Engineer and currently is working toward the Ph.D. degree at Purdue University, West Lafayette IN. His main research interests are in the modeling and simulation of quantum transports with a secondary focus on the performance optimization in the nano-electronic computations.

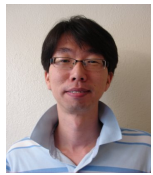

**Insoo Woo** received the BS degree in Computer Engineering in 1998 from Dong-A University in South Korea and was employed as a software engineer during 1997 to 2006. He is a PhD student in School of Electrical and Computer Engineering at Purdue University. His research interest is GPU-Aided Techniques for Computer Graphics and Visualization.

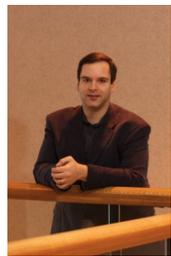

**David Ebert** is a Professor in the School of Electrical and Computer Engineering at Purdue University, a University Faculty Scholar, a Fellow of the IEEE, Director of the Purdue University Rendering and Perceptualization Lab (PURPL), and Director of the Purdue University Regional Visualization and Analytics Center (PURVAC), which is part of the Department of Homeland Security's Regional Visualization and Analytics Center of Excellence. Dr. Ebert performs research in novel visualization techniques, visual analytics, volume rendering, information visualization, perceptually-based visualization, illustrative visualization, and procedural abstraction of complex, massive data. Ebert has been very active in the visualization community, teaching courses, presenting papers, co-chairing many conference program committees, serving on the ACM SIGGRAPH Executive Committee, serving as Editor in Chief of IEEE Transactions on Visualization and Computer Graphics, serving as a member of the IEEE Computer Society's Publications Board, serving on the National Visualization and Analytics Center's National Research Agenda Panel, and successfully managing a large program in external funding to develop more effective methods for visually communicating information.

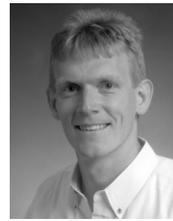

**Gerhard Klimeck** is the Associate Director for Technology of the Network for Computational Nanotechnology and Professor of Electrical and Computer Engineering at Purdue University. He leads the development and deployment of web-based simulation tools that are hosted on http://nanohub.org a community website that is utilized by over 87,000 users annually. Over 6,700 users have run over 375,000 simulations in the past 12 months. He was the Technical Group Supervisor for the Applied Cluster Computing Technologies Group and continues to hold his appointment as a Principal Member at the NASA Jet Propulsion Laboratory on a faculty part-time basis. Previously he was a member of technical staff at the Central Research Lab of Texas Instruments. He received his Ph.D. in 1994 from Purdue University and his German electrical engineering degree in 1990 from Ruhr-University Bochum. His research interest is in the modeling of Nanoelectronic devices, parallel cluster computing, and genetic algorithms. He has been the lead on the development of NEMO 3-D, a tool that enables the simulation of tens-of-million atom quantum dot systems, and NEMO 1-D, the first Nanoelectronic CAD tool. Dr. Klimeck's work is documented in over 220 peer-reviewed publications and over 360 conference presentations. He is a member of IEEE (senior), APS, HKN and TBP. More information can be found at http://www.ece.purdue.edu/~gekco and http;//nanoHUB.org/klimeck .